\documentclass[prd,aps,preprint,tightenlines]{revtex4}
\usepackage{epsfig}
\usepackage{amsmath}
\vspace{1in}
\usepackage{axodraw}
\begin{document}
\title{Scale Dependence of Hadronic Wave Functions \\and Parton Densities}
\author{Matthias Burkardt\footnote{Permenant address: Department of Physics,
New Mexico State University, Las Cruces, NM 88003}}
\affiliation{Department of Physics,
University of Maryland,
College Park, Maryland 20742 }
\author{Xiangdong Ji}
\email{xji@physics.umd.edu}
\affiliation{Department of Physics,
University of Maryland,
College Park, Maryland 20742 }
\author{Feng Yuan}
\email{fyuan@physics.umd.edu}
\affiliation{Department of Physics,
University of Maryland,
College Park, Maryland 20742 }
\date{\today}          
\begin{abstract}

We study how the components of hadronic wave functions in 
light-cone quantization depend on the ultraviolet cut-off
by relating them in a systematic way to the matrix elements
of a class of quark-gluon operators between the QCD vacuum
and the hadrons. From this, we derive an infinite set of 
scale-evolution equations for the individual contributions 
to parton distributions from the Fock expansion. When 
summed over all the contributions, we recover the well-known 
DGLAP equation.  

\vspace{10cm}
\end{abstract}
\maketitle
\newcommand{\be}{\begin{equation}}
\newcommand{\ee}{\end{equation}}
\newcommand{\ben}{\[}
\newcommand{\een}{\]}
\newcommand{\beqn}{\begin{eqnarray}}
\newcommand{\eeqn}{\end{eqnarray}}
\newcommand{\Tr}{{\rm Tr} }

In light-cone quantization and light-cone gauge, 
the hadronic states in QCD are expressed as an expansion of 
various quark and gluon Fock components \cite{brodsky}. This 
expansion depends, among others, on the momentum cut-off used to 
truncate the theory, which is often interpreted  
as the physical resolution scale. Although the physical
observables, such as masses, angular momenta, form factors
and cross sections, ought be independent of the cut-off, 
many interesting hadronic matrix elements do. A well-known example
is the matrix elements of twist-two operators which define the moments
of Feynman's parton distributions \cite{text}. For these quantities,
one should be able to trace the scale dependence back to that of the 
hadronic wave functions. 

In this paper, we are interested in 
how the light-cone Fock components depend on the momentum cut-off. 
Finding the solution directly from 
diagonalizing the light-cone hamiltonian is less obvious. Instead
we approach the problem by systematically relating
the Fock expansion to the matrix elements of a certain class
of quark-gluon operators between the QCD vacuum $|0\rangle$ and the 
hadron states, taking advantage of the simplicity of 
$|0\rangle$ in light-cone quantization. The scale-dependence 
of the wave function amplitudes can then be traced to the wave 
function renormalization constants of quark and gluon fields. 
Following this, we derive the scale dependence of the parton 
densities from individual Fock components. 
The scale evolution of these contributions
obeys an infinite set of coupled, linear differential-integral 
quations. When summed over 
all Fock contributions, we recover the well-known Dokshitzer-Gribov-
Lipatov-Altarelli-Parisi (DGLAP) equation for parton densities.

Before starting, let us remind the reader some salient features
of light-cone quantization relevant for the following discussion \cite{bl}.
The light-cone time $x^+$ and coordinate $x^-$ are defined as $x^\pm
=1/\sqrt{2}(x^0\pm x^3)$. Likewise we define Dirac matrices
$\gamma^\pm = 1/\sqrt{2}(\gamma^0\pm\gamma^3)$. The 
projection operators for Dirac fields are defined 
as $P_\pm = (1/2)\gamma^\mp\gamma^\pm$. Any Dirac field $\psi$ 
can be decomposed into $\psi=\psi_++\psi_-$ with
$\psi_\pm = P_\pm \psi$. $\psi_+$ is a dynamical degrees of freedom 
and has the canonical expansion, 
\begin{eqnarray}
  \psi_+(\xi^+=0,\xi^-,\xi_\perp)
  &=& \int {d^2k_\perp\over (2\pi)^3}
    {dk^+\over 2k^+}
   \sum_\lambda
  \left[b_\lambda(k) u(k\lambda) e^{-i(k^+\xi^--\vec{k}_\perp\vec{\xi}_\perp)} \right. \nonumber \\
 && \left. + d_\lambda^\dagger(k) v(k\lambda)e^{i(k^+\xi^--\vec{k}_\perp\vec{\xi}_\perp)}
 \right]
  \ . 
\end{eqnarray}
Likewise, for the gluon fields in the light-cone gauge $A^+=0$, 
$A_\perp$ is dynamical and has the expansion,
\begin{eqnarray}
    A_\perp(\xi^+=0,\xi^-,\xi_\perp)
  &=&  \int {d^2k_\perp\over (2\pi)^3}
    {dk^+\over 2k^+}
   \sum_\lambda
  \left[ a_\lambda(k) \epsilon(k\lambda) e^{-i(k^+\xi^--\vec{k}_\perp\vec{\xi}_\perp)}
   \right.
  \nonumber \\
  && \left. + a_\lambda^\dagger(k) \epsilon^*(k\lambda)e^{i(k^+\xi^--\vec{k}_\perp
   \vec{\xi}_\perp)} \right] \ . 
\end{eqnarray}
$\psi_-$ and $A^-$ are dependent variables.

The key observation in this paper is that the light-cone Fock 
expansion of a hadron state is {\it completely} defined by the 
matrix elements of a special class equal light-cone time 
quark-gluon operators between the QCD vacuum and the hadron. 
These operators are specified as follows: Take the + component
of the Dirac field $\psi_+$ and the $+\perp$-component of the
gauge field $F^{+\perp}$. [We sometimes label the $\perp$ components
with index $i=1, 2$.] Assume all these fields are at light-cone
time $x^+=0$, but otherwise with arbitrary dependence on other spacetime 
coordinates. Products of these fields with the right 
quantum numbers (spin, flavor, and color) define a set of  
operator basis. [This has been done in the
past for light-cone correlations in which all fields are separated along the
light-cone \cite{lcc}.] Clearly, these operators are not gauge-invariant 
because one cannot gauge-invariantize them by simply inserting
string operators along the light-cone. Since all fields are  
at different points in the transverse directions, the operators do not have
singularities requiring special renormalization.
Moreover, at equal light-cone time, there is no need to 
introduce a time-ordering among different fields because
the difference is proportional to equal-light-cone-time commutators 
which are straightfoward to evaluate. In fact, to simplify the discussion, 
we assume all fields in the operators are normal
ordered, i.e., the annihilation operators appearing at 
the right of the creations. We believe that the matrix elements of all these 
operators between the hadron state and the QCD vacuum yield {\it complete} 
information about the hadron wave function. 

As an example, let us consider $\pi^+$ meson with
momentum $P^\mu$ along the $z$-direction. The leading light-cone 
Fock states consist of a pair of up and anti-down quarks. 
The light-cone helicity of the $\pi^+$ meson is 
zero, but the light-cone helicity of the
quark-antiquark pair can either be zero or $\pm 1$.
Use $u_+(\xi^-,\xi_\perp)$ to represent the up-quark
field in the coordinate space and ${\overline d}_+(0)$ the
anti-down-quark field. In the massless limit, the operator
$\overline{d}_+\gamma^+\gamma_5 u_+$
yields a helicity-0 pair, and $\overline{d}_+\sigma^{+\perp}
\gamma_5 u_+$ a helicity-1 pair. The helicity counting here is based
on the chirality of the operators and the relation between 
chirality and helicity of massless fermions. [One can in principle
construct similar operators without the $\gamma_5$ matrix; however
parity forbids any finite matrix elements of them between the QCD vacuum and  
pseudo-scalar mesons.] The first operator defines a coordinate 
amplitude,
\begin{equation}
   \langle 0|\overline{d}_+(0)\gamma^+
  \gamma_5 u_+(\xi^-,\xi_\perp)|\pi^+(P)\rangle
   = \phi_0(\xi^-, \xi_\perp)2
P^+ \ , 
\label{amp1}
\end{equation}
where we normalize the state covariantly $\langle P|P'\rangle
=2P^+(2\pi)^3\delta(P^+-{P^+}')\delta^2(P_\perp-P_\perp')$. 
Introducing the Fourier tranformation of the amplitude
\begin{equation}
    \phi_0(k_\perp, x) = \int d^2\xi_\perp d\xi^- e^{i(k^+\xi^--k_\perp\xi_\perp)} 
 \phi_0(\xi^-, \xi_\perp) \ , 
\end{equation}
we can invert Eq. (\ref{amp1}) to find
\begin{eqnarray}
  |\pi^+(P)\rangle &=& \int {d^2k_\perp\over (2\pi)^3}
    {dx\over 2\sqrt{x(1-x)}} \phi_0(x,k_\perp)  
\left[b^\dagger_{u\uparrow i}(x,\vec{k}_\perp) 
  d^\dagger_{u\downarrow i}(1-x,-\vec{k}_\perp) \right. \nonumber \\
&& - \left. b^\dagger_{u\downarrow i}(x,\vec{k}_\perp)
  d^\dagger_{d\uparrow i}(1-x,-\vec{k}_\perp)\right]
 |0\rangle +...
\end{eqnarray}
where $i$ is the color index. The creation and annihilation operators
are normalized according to the commutation relation $[b(k),b^\dagger(k')]_+
= 2k^+(2\pi)^3\delta(k^+-{k^+}')\delta^2(k_\perp-k_\perp')$. 
We have assumed here that the full QCD vacuum is a perturbative vacuum 
in light-cone quantization. In particular, we neglect
the subtlety of zero-modes which might cause problems 
at some stage \cite{bh}.

The operator with helicity-1 quark-anti-quark pair defines the amplitude
\begin{equation}
   \langle 0|\overline{d}_+(0)\sigma^{+i}
  \gamma_5 u_+(\xi^-,\xi_\perp)|\pi^+(P)\rangle
   = \partial^i\phi_1(\xi^-, \xi_\perp) 2P^+ \ , 
\label{amp2}
\end{equation} 
where $i=1,2$ is an index for transverse directions.
Peforming a Fourier transformation on the both sides and inverting the
equation, we find a light-cone Fock component
\begin{eqnarray}
  |\pi^+(p)\rangle &=& \int {d^2k_\perp\over (2\pi)^3}
    {dx\over 2\sqrt{x(1-x)}}
   \phi_1(x,k_\perp) \left[ (k_1-ik_2)
   b^\dagger_{u\uparrow i}(x,\vec{k}_\perp) d^\dagger_{d\uparrow i}
 (1-x,-\vec{k}_\perp)
\right. \nonumber \\
&& + \left. (k_1+ik_2)b^\dagger_{u\downarrow i}(x,\vec{k}_\perp)
d^\dagger_{d\downarrow i}(1-x,-\vec{k}_\perp) \right] |0\rangle +...
\end{eqnarray}
The angular momentum content of the wave function is clear:
For a quark-antiquark pair carrying helicity $\pm 1$, it couples
to an orbital wave function with $L_z=\mp 1$. Parity 
determines the relative sign of the two contributions. The phenomenological
implications of $\phi_1(x, k_\perp)$ for pion form factors have
been discussed in the literature before \cite{pff}. 

It is now straightfoward to study the cut-off dependence of 
$\phi_{0,1}(x,k_\perp)$.
For the moment, we focus on the transverse momentum 
cut-off $\Lambda$, although a cut-off in $x$ is also 
needed at $x\rightarrow 0$ in general. 
Besides the explicit cut-off dependence in the wave function
amplitudes, the $k_\perp$ integration 
in Eq. (5) is implicitly bounded by $\Lambda$. 
In any cut-off scheme, the quark and gluon fields in QCD as well 
as the strong coupling constant $\alpha_s$ depend on the cut-off. 
For large $\Lambda$, the dependence of quantum fields on $\Lambda$
can be calculated in perturbation theory because of the asymptotic 
freedom. In fact, according to 
the standard renormalization theory, on has
\begin{equation}
     \psi_\Lambda (\xi) = Z_F^{1/2}(\Lambda) \tilde \psi(\xi), ~~~
      A_\lambda^\mu(\xi) = Z_A^{1/2}(\Lambda) \tilde A^\mu(\xi)
\end{equation}
where $\tilde \psi(\xi)$ and $\tilde A^\mu(\xi)$ are 
indepedent of $\Lambda$,  and $Z_{F, A}(\Lambda)$ 
are the wave function renormalization constants. Although 
the factorization in the above equation is scheme-dependent, 
but the $\Lambda$ dependence itself is not.

Going back to Eqs. (\ref{amp1},\ref{amp2}),  it is now clear that
the cut-off dependence of the wave function amplitudes
$\phi_i$ comes entirely from the field renormalization,
\begin{equation}
      \phi^\Lambda_{i}(x, k_\perp) = Z_F(\Lambda) 
       \tilde \phi_i(x, k_\perp) \ , 
\end{equation}
where $\tilde \phi_i(x, k_\perp)$ is independent of $\Lambda$. 
 
We claim that the above feature holds for all 
components of the pion wave function in the Fock expansion.
For instance, the most general two-quark, one-gluon 
wave function amplitudes can be defined through the matrix 
elements of the operators, 
\begin{eqnarray}
   &&  \overline{d}_+(0)\gamma^+\gamma_5 
F^{+j}(\eta_-, \eta_\perp) u_+(\xi^-,\xi_\perp) \ , \nonumber \\
   && \overline{d}_+(0)\sigma^{+i}\gamma_5 
F^{+j}(\eta_-, \eta_\perp) u_+(\xi^-,\xi_\perp) \ . 
\end{eqnarray}
Their scale dependence comes entirely from the wave function 
renormalization constants $Z_F(\Lambda)Z_A^{1/2}(\Lambda)$. 
In general, an $n$-particle Fock wave function amplitude 
with $n_q$ quark and antiquark and $n_g$ gluon creation operators
has an explicit cut-off dependence through $Z_F^{n_q/2}(\Lambda)
Z_A^{n_g/2}(\Lambda)$. Once again, all the momentum integrations
are cut-off by $\Lambda$.  

Knowing the scale dependence of individual components of the
hadron wave function, we can calcualte the scale dependence of 
their contributions to hadronic matrix elements. As an example, 
we consider in the remainder of the paper Feynman parton 
distributions, although the discussion applies to generalized
parton distributions as well \cite{ji}.  
Deriving the parton evolution equation from light-cone wave functions 
has been considered in Ref. \cite{bhl}. Our approach here allows 
to uncover a set of new equations.  

Consider, for example, the two-particle wave-function 
contribution to the $u$ quark 
distribution in the pion. We have
\begin{equation}
     u_2(x,\Lambda) = \int^\Lambda_0 
  {d^2k_\perp\over (2\pi)^3} 
       \left[|\phi^\Lambda_0(x, k_\perp)|^2 + k_\perp^2|\phi_1^\Lambda(x,
   k_\perp)|^2 \right]\ . 
\end{equation}
Since at large $k_\perp$, $\phi_0(x,k_\perp)$ goes
like $1/k_\perp^2$ and $\phi_1(x,k_\perp)$ like $1/k_\perp^4$
modulo logaritms \cite{brodsky}, the $k_\perp$
integration is convergent and hence $\Lambda$ can be
taken to infinity. Thus the only $\Lambda$ dependence
in $u_2(x, \Lambda)$ comes from the wave function renormalization factor
$Z_F$. This yields the 
following evolution equation for $u_2(x, \Lambda)$
\begin{equation}
     {d\over d\ln \Lambda^2} u_2(x,\Lambda)
   = -2\gamma_F {\alpha_s(\Lambda)\over 4\pi} 
     u_2 (x, \Lambda) \ , 
\end{equation}
where $\gamma_F$ is the anomalous dimension 
of $Z_F$ and is gauge-dependent. [In physical gauges, 
it is positive-definite.]
In light-cone gauge,
\begin{equation}
       \gamma_F
    = 2C_F\int^1_0 dy {1+(1-y)^2\over y} \ , 
\end{equation}
where $C_F=4/3$. The above integral diverges at $y=0$ \cite{brol}, 
and we regulate 
the integral by cutting it off at $y=\epsilon$. Physics
of course must be independent of any cut-off. 
 
Equation (12) indicates that the two-particle Fock state contribution to 
the up-quark distribution graduately diminishes
as $\Lambda\rightarrow \infty$. The physics  
is simple: As $\Lambda$ gets larger, the
state is probed is at shorter distance, and it becomes
increasingly difficult for the meson to remain
in the two-particle Fock component because of the radiation. 
As a consequence, the three-particle Fock amplitude 
increases at the leading-logarithmic level. 
In the light-cone gauge, $\gamma_F$ diverges
at small $x$, and the radiation rate will depend on the
cut off $\epsilon$. 

Consider now the three-particle Fock component
contribution to the $u$-quark distribution. We write schematically, 
\begin{equation}
       u_3 (x, \Lambda) =  \int {d^2k_\perp \over (2\pi)^3}
    {d^2k_\perp' dx'\over (2\pi)^3 }
  |\phi^\Lambda(x,k_\perp, x', k_\perp', \Lambda)|^2\ , 
\end{equation}
where we have not considered individual quark-antiquark-gluon
helicities and orbital angular momentum projections although
this can be done straightforwardly.  
As discussed before, the wave function
$\phi^\Lambda (x, k_\perp, x', k_\perp')$ has an explicit
dependence on $\Lambda$ through the wave
function renormalization constant $Z_FZ_A^{1/2}$. 
Additional dependence comes from integrations over
the transverse momenta $k_\perp$ and $k_\perp'$. 

We take derivative with respect to $\Lambda$ using 
the chain rule. The derivative with respect
to the wave function renormalization yields, 
\begin{equation}
     {d\over d\ln \Lambda^2} u_3(x,\Lambda)
   = -(2\gamma_F +\gamma_A) {\alpha_s(\Lambda)\over 4\pi} 
     u_3 (x, \Lambda) +...\ . 
\end{equation}
where $\gamma_A$ is the anomalous dimension
of the gluon wave function renormalization. The physics
of this part of the scale evolution is the same
as the two-particle Fock component case: The splitting of the
partons leads to the decrease of the probability 
for the pion to remain in the three-particle Fock state. 

The integrations over $\vec{k}_\perp$ and 
$\vec{k}_\perp'$ do not yield divergences in general. 
In fact, there is no overall divergence (divergence arising from 
when all transverse momenta going to infinity at the same rate) 
because the power counting indicates that when two-momenta 
going to infinity at the same time, 
the integrals have negative superficial degree of 
divergence. However, there are subdivergences. 
These subdivergences arise from one-loop diagrams
shown in Fig 1. The physics of these diagrams 
is that there are three-particle Fock amplitudes
which are generated from the radiation of
the two-particle Fock amplitude. Therefore, the result
is proportional to $u_2(x, \Lambda)$. 

\begin{figure}
\SetWidth{0.7}
\begin{center} \begin{picture}(200,100)(0,0)
\Line(20,10)(200,10)
\LongArrow(60,10)(55,10)
\Text(55,15)[b]{\Large $\bar d$}
\Text(55,68)[]{\Large $u$}
\LongArrow(165,10)(160,10)
\GlueArc(110,10)(25,0,180){3}{6}
\SetWidth{4}
\ArrowLine(-20,30)(20,30)
\ArrowLine(204,30)(240,30)
\SetWidth{0.7}
\ArrowLine(20,50)(100,70)
\ArrowLine(120,70)(200,50)
\GOval(20,30)(20,4)(0){0.9}
\GOval(200,30)(20,4)(0){0.9}
\CArc(110,70)(10,0,360)
\Text(110,70)[]{\LARGE X}
\DashLine(110,90)(110,0){5}
\Text(110,-15)[]{\Large (a)}
\end{picture}  
\end{center}

\vskip 1cm

\begin{center} \begin{picture}(200,100)(0,0)
\Line(20,10)(200,10)
\GlueArc(110,70)(35,192,348){3}{7}
\Text(55,15)[b]{\Large $\bar d$}
\Text(55,68)[]{\Large $u$}
\SetWidth{4}
\ArrowLine(-20,30)(20,30)
\ArrowLine(204,30)(240,30)
\SetWidth{0.7}
\LongArrow(60,10)(55,10)
\LongArrow(165,10)(160,10)
\ArrowLine(20,50)(100,70)
\ArrowLine(120,70)(200,50)
\GOval(20,30)(20,4)(0){0.9}
\GOval(200,30)(20,4)(0){0.9}
\CArc(110,70)(10,0,360)
\Text(110,70)[]{\LARGE X}
\DashLine(110,90)(110,0){5}
\Text(110,-15)[]{\Large (b)}
\end{picture}  
\end{center}
\caption{Contributions to the u-quark distribution from the three-particle
component generated by the two-particle Fock component.}
\end{figure}
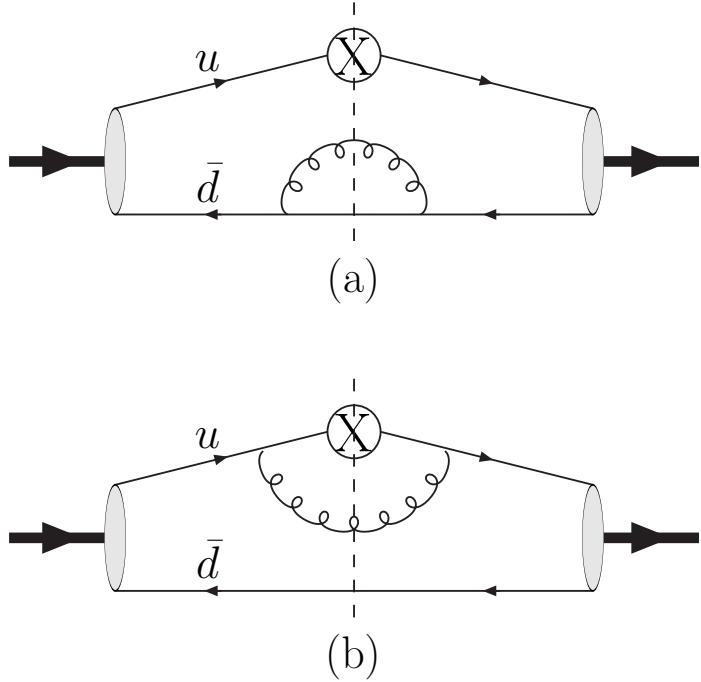

First consider the loop intergral from the 
$d$-quark line shown in Fig. 1a. In this case, the $\vec{k}_\perp'$ 
integration is divergent whereas the $\vec{k}_\perp$ integration 
is finite. It is easy to see that when $\Lambda$ changes, 
the result 3-particle distributions
also changes.  From this diagram, one finds,
\begin{equation}
     {d\over d\ln \Lambda^2} u_3(x,\Lambda)  
 = ... + \gamma_F {\alpha_s(\Lambda)\over 4\pi} u_2(x, \Lambda) \ . 
\end{equation}
where the plus sign indicates that the three-particle Fock 
component receives a contribution from the radiation of 
the two-parton states. 

Finally, let us consider the gluon radition from
for the u-quark line as shown in Fig. 1b. The integration over
$\vec{k}_\perp$ is now divergent whereas the one over  $\vec{k}_\perp'$ 
is finite. The $\vec{k}_\perp$ integration 
can be done using the standard light-cone
perturbation theory,
\begin{equation}
      {d\over d\ln \Lambda^2} u_3(x,\Lambda)  
 = ... + {\alpha_s(\Lambda)\over 2\pi} C_F
   \int^1_x {dy\over y} {1+y^2\over 1-y}
     u_2\left({x\over y}, \Lambda\right) \ ,  
\end{equation}
where the divergence at $y=1$ must be regulated. 
Adding everything together, we obtain the
complete evolution equation for $u_3$ is
\begin{eqnarray}
{d\over d\ln \Lambda^2} u_3(x,\Lambda)  
 &=& \left.{\alpha_s(\Lambda)\over 4\pi} \right[-(2\gamma_F+\gamma_A) 
  u_3(x,\Lambda) + \gamma_F u_2(x,\Lambda)
      \nonumber \\
  && \left. +  2C_F\int^1_x {dy\over y} {1+y^2\over 1-y}
     u_2\left({x\over y}, \Lambda\right) \right] \ . 
\end{eqnarray}
This is an inhomogeneous equation with a 
driving term $u_2$. 

Going to wave function amplitudes with four and more partons
posts no special difficulty, except one has to take into account 
the mixing with the singlet contribution. Take the example of 
four-parton amplitudes for which three flavor structures $u\overline{d}gg$, 
$u\overline{d}\overline{u}u$, and $u\overline{d}\overline{d}d$ must be 
considered separately. 
For $u\overline{d}gg$, the gluon radition from $u$, $\overline{d}$ an $g$ of the
three-parton component $u\overline{d}g$ yields, 
\begin{eqnarray}
{d\over d\ln \Lambda^2} u_4^{u\overline{d}gg}(x,\Lambda)  
 = && {\alpha_s(\Lambda)\over 4\pi} \left[-2(\gamma_F+\gamma_A) 
  u_4^{u\overline{d}gg}(x,\Lambda) \right. \nonumber  \\
  &&  \left. + (\gamma_F+\gamma_{A1}) u_3(x,\Lambda)
  +   2C_F\int^1_x {dy\over y} {1+y^2\over 1-y}
     u_3\left({x\over y}, \Lambda\right) \right] \ . 
\end{eqnarray}
where $\gamma_{A1}$ is the part of the gluon anomalous dimension
from the gluon loop. 
On the other hand, for $u\overline{d}\overline{u}u$, the gluon splitting into $u\overline{u}$
pair yields, 
\begin{eqnarray}
{d\over d\ln \Lambda^2} u_4^{u\overline{d}u\overline{u}}(x,\Lambda)  
 &=& {\alpha_s(\Lambda)\over 4\pi} \left[-4\gamma_F 
  u_4^{u\overline{d} u\overline{u}}(x,\Lambda) + \gamma_{A2} u_3(x,\Lambda)
   \right. \nonumber \\
 && \left. +   2\int^1_x {dy\over y} T_F(y^2+(1-y)^2)
     g_3\left({x\over y}, \Lambda\right) \right] \ , 
\end{eqnarray}
where $T_F=1/2$ and $\gamma_{A2}$ is the part of the gluon anomalous dimension
from the ${\overline u}u$ loop. $g_3(x,\Lambda)$ is the gluon distribution
from the $u\overline{d}g$ Fock amplitude. 
Finally, for $ud\overline{d}d$, the gluon splitting into $d\overline{d}$
pair yields, 
\begin{equation}
 {d\over d\ln \Lambda^2} u_4^{u\overline{d}d\overline{d}}(x,\Lambda)  
 = {\alpha_s(\Lambda)\over 4\pi} \left[-4\gamma_F 
  u_4^{u\overline{d} d\overline{d}}(x,\Lambda) + \gamma_{A3} u_3(x,\Lambda)
   \right] \ . 
\end{equation}
where $\gamma_{A3}$ is the part of the gluon anomalous dimension
from the ${\overline d}d$ loop. 
When adding all the contributions $(\gamma_A=\gamma_{A1}+\gamma_{A2}
+\gamma_{A3}$) , we have, 
\begin{eqnarray}
 {d\over d\ln \Lambda^2} u_4(x,\Lambda)  
 = && \left.{\alpha_s(\Lambda)\over 4\pi} \right[-4\gamma_F 
  u_4^{u\overline{d} q\overline{q}}(x,\Lambda) 
  - 2(\gamma_F+\gamma_A) u_4^{u\overline{d}gg}(x, \Lambda) 
   + (\gamma_F+\gamma_A) u_3(x,\Lambda) \nonumber \\
&&
\left. + 2 \int^1_x {dy\over y} \left\{ C_A{1+y^2\over 1-y} 
     u_3\left({x\over y}, \Lambda\right) + T_F
(y^2+(1-y)^2) g_3\left({x\over y}, \Lambda\right) \right\} 
   \right] \ . 
\end{eqnarray}

To keep the evolution simple, we also consider the anti-up quark
distribution at this order. The only contribution is from $u\overline{d}u\overline{u}$
component for which we have 
\begin{equation}
    {d\over d\ln \Lambda^2} \overline{u}_4(x,\Lambda)  
 = {\alpha_s(\Lambda)\over 4\pi} \left[-4\gamma_F 
  \overline{u}_4(x,\Lambda) 
  +  2\int^1_x {dy\over y} T_F(y^2+(1-y)^2)
     g_3\left({x\over y}, \Lambda\right) 
   \right] \ . 
\end{equation}
Therefore, if we define the valence up-quark distribution
$u_v=u-\overline{u}$, then
\begin{eqnarray}
 {d\over d\ln \Lambda^2} u_{4v}(x,\Lambda)  
 &=& {\alpha_s(\Lambda)\over 4\pi} \left[-4\gamma_F 
  u_{4v}^{u\overline{d} q\overline{q}}(x,\Lambda) 
  - 2(\gamma_F+\gamma_A) u_{4v}^{u\overline{d}gg}(x, \Lambda) \right. \nonumber \\
&& \left. + (\gamma_F+\gamma_A) u_3(x,\Lambda)
+  2\int^1_x {dy\over y} {1+y^2\over 1-y} u_3\left({x\over y}, \Lambda\right) 
   \right] \ . 
\end{eqnarray}
without the complication from the $g\rightarrow q\overline{q}$ kernel. 
 
It is not difficult to see that the evolution equation
for $u_{nv}$ from Fock states with $n$ partons is 
\begin{eqnarray} 
{d\over d\ln \Lambda^2} u_{nv}(x,\Lambda)  
 &=& \left. {\alpha_s(\Lambda)\over 4\pi} \right[-\sum_{i=1}^n\gamma_i 
  u_{nv}(x,\Lambda) + \sum_{i=1}^{n-2} \gamma_i u_{n-1v}(x, \Lambda)
  +  \nonumber \\
 &&  \left.  2C_F\int^1_x {dy\over y} {1+y^2\over 1-y}
     u_{n-1v}({x\over y}, \Lambda) \right] \ , 
\end{eqnarray}
where the sum over $n-2$ $\gamma_i$ excludes one $\gamma_F$ and one $\gamma_A$. 
The first term in Eq. (25) describes the depletion of the $n$-particle
Fock component due to the gluon radiation into $n+1$-particle component; 
the second and third terms describe the increase of the $n$-particle component
due to the gluon emission of the $n-1$-particle component. The difference
between the later two comes from whether the gluon is radiated from the
active particle or the spectators. 
The total $u_v(x)$ distribution is a sum over all possible Fock components,
\begin{equation}
          u_v(x) = \sum_{i=2}^\infty u_{iv}(x) \ . 
\end{equation}
Summing over all the equations for the individual Fock components, we 
recover the standard DGLAP equation,
\begin{equation}
{d\over d\ln \Lambda^2} u_v(x,\Lambda)  
 = {\alpha_s(\Lambda)\over 2\pi}  
   \int^1_x {dy\over y} {1+y^2\over (1-y)_+}
     u_v\left({x\over y}, \Lambda\right)
\end{equation}
which is an important check.
Using the same procedure, one can derive evolution
equations for other types of parton distributions, 
such as the singlet quark and gluon distributions,
quark helicity and transversity distributions, as
well as higher-twist distributions.

In summary, we find that the light-cone wave functions of hadrons
in QCD can be entirely determined by the matrix elements of a class
of quark-gluon operators. From this, we derive an infinite set of
evolution equations for the parton distributions contributed
by individual $n$-particle Fock components. These equations are consistent
with the well-known DGLAP equation and are useful for phenomenological
studies of hadronic structures and model buildings. They should also 
provide important constraints for wave functions derived from light-cone
quantization. 

Note added: After this paper was finished, we were
informed that Ref. \cite{hari} has studied the
evolution of the 2 and 3 particle Fock components,
using an explicit light-front calculation.

The authors thank S. Brodsky for a number of useful discussions and
encouragement. X. J. acknowledges a conversation with F. Low 
about light-cone wave functions many years ago. 
This work was supported by the U. S. Department of Energy via 
grants FG-03-95ER-40965 and DE-FG02-93ER-40762.


\begin{references}
\frenchspacing

\bibitem{brodsky}
See for example, S. J. Brodsky, Invited Lectures presented at the Cracow School
of Theoretical Physics, Zakopane, Poland, June 2001, hep-ph/0111340. 

\bibitem{text}
See for example, M. E. Peskin and D. V. Schroeder, 
{\it An Introduction to Quantum Field Theory}, Addison-Wesley, New York, 1995.  

\bibitem{bl}
See for example, J. B. Kogut and D. E. Soper, Phys.\
Rev.\ D {\bf 1}, 2901 (1971). 
S. Brodsky and P. Lepage, {\it Introduction to Perturbative QCD}, 
Ed. by A. Mueller, World Scientific, Singapore, 
1989; P.~P.~Srivastava and S.~J.~Brodsky,
Phys.\ Rev.\ D {\bf 64}, 045006 (2001).

\bibitem{lcc}
V. M. Braun, S. E. Derkachov, G. P. Korchemsky and A. N. Manashov,
Nucl. Phys. B553, 355 (1999); P. Ball, V. M. Braun, Y. Koike, 
and K. Yanaka, Nucl. Phys. B529, 323 (1998). 

\bibitem{bh}
S. J. Brodsky, P. Hoyer, N. Marchal, S. Peigne, F. Sannino, hep-ph/0104291. 

\bibitem{pff}
N. Isgur and C. H. Llewellyn-Smith, Nucl. Phys. B317, 526 (1989); 
S. W. Wang and L. S. Kisslinger, Phys. Rev. D54, 5890 (1996); 
F. G. Cao, J. Cao, T. Huang, and B. Q. Ma, Phys. Rev. D55, 7107 (1997). 

\bibitem{ji}
X. Ji, J. Phys. G 24, 1181 (1998);
A. V. Radyushkin, hep-ph/0101225; K. Goeke, M. V. Polyakov, 
M. Vanderhaeghen, Prog. Part. Nucl. Phys. 47, 401 (2001). 

\bibitem{bhl}
S. J. Brodsky, T. Huang, and G. P. Lepage, 
talk presented at the Summer Institute on Particle Physics, 
SLAC, Aug. 1981, SLAC-PUB-2857. 

\bibitem{brol}
G. P. Lepage and S. J. Brodsky, 
Phys.\ Lett.\ B {\bf 87}, 359 (1979);
Phys.\ Rev.\ D {\bf 22}, 2157 (1980). 

\bibitem{hari}
A. Harindranath, R. Kundu, and W.-M. Zhang,
Phys. Rev. D59, 094013 (1999).

\end{references}
\end{document}